**Title**

Near-IR wide field-of-view Huygens metalens for outdoor imaging applications.

**Authors**

J. Engelberg,[1] C. Zhou,[2] N. Mazurski,[1] J. Bar-David,[1] A. Kristensen,[2,3] and U. Levy,[1]*

**Affiliations**

[1] Department of Applied Physics, The Faculty of Science, The Center for nanoscience and nanotechnology, The Hebrew University Jerusalem Israel, Jerusalem, Israel, 91904
[2] Department of Micro- and Nanotechnology, Technical University of Denmark, DK-2800 Kongens Lyngby, Denmark
[3] Department of Health Technology, Technical University of Denmark, DK-2800 Kongens Lyngby, Denmark

* ulevy@mail.huji.ac.il

**Abstract**

The ongoing effort to implement compact and cheap optical systems is the main driving force for the recent flourishing research in the field of optical metalenses. Metalenses are a type of metasurface, used for focusing and imaging applications, and are implemented based on the nanopatterning of an optical surface. The challenge faced by metalens research is to reach high levels of performance, using simple fabrication methods suitable for mass-production. In this paper we present a Huygens nanoantenna based metalens, designed for outdoor photographic/surveillance applications in the near-infra-red. We show that good imaging quality can be obtained over a field-of-view (FOV) as large as ±15˚. This first successful implementation of metalenses for outdoor imaging applications is expected to provide insight and inspiration for future metalens imaging applications.

**MAIN TEXT**

**Introduction**

The topic of metasurfaces in general, and the specific case of metalenses, is an ongoing area of research (*1–4*). The main aspirations for metalenses are miniaturization and cost reduction of optical systems by replacement of conventional lenses with metalenses.

Most metalenses operate as a first order diffractive lens, i.e. the lens introduces a phase function modulo $2\pi$ to the wavefront, that ideally converts one spherical wavefront (emanating from a point source) to another (by focusing down to a diffraction limited spot). The difference between a metalens and the more traditional surface relief diffractive lens (*5*) is that in a surface relief diffractive lens the phase is introduced via optical path difference (OPD) resulting from the physical profile of the substrate, while in a metalens the phase is affected via nanostructures patterned on the substrate, that introduce phase delay. Three common methods for introducing the phase delay in dielectric metasurfaces are truncated waveguides (*6*), geometrical phase (*7*), and Huygens nanoantennas (*8*), each having their pros and cons.

Truncated waveguide metasurfaces are composed of nanorods, whose diameters determine the effective refractive index of the optical mode propagating through them, thus allowing phase control. Geometrical phase metasurfaces consist of nanofins, whose orientation determines the phase shift imparted to the light. Huygens metasurfaces consists of resonant structures, e.g. nanodisks. The nanodisks are designed so that the electric and magnetic dipole resonances overlap (thus meeting what is known as the first Kerker condition in which reflection is inhibited by destructive interference), which results in excellent transmission, and a $2\pi$ phase shift range. The phase shift can be controlled by changing the diameter of the disk (*1*).

The advantage of a Huygens metalens over the other two types is the low aspect-ratio (height to diameter/width) of the nanoantennas, which makes it easier to manufacture. This comes at the expense of higher sensitivity to wavelength and incidence angle. The purpose of this paper is to explore what performance can be achieved by a Huygens metalens in the optical region, over different wavelength ranges and fields-of-view, and to provide the first ever demonstration of a Huygens type metalens for outdoor imaging applications.

Huygens metalenses have been demonstrated in the microwave spectral region (*9*, *10*), and recently also in the optical spectral region (*11*, *12*). These optical region metalenses are designed for monochromatic operation, high-NA, and narrow field-of-view. A Wide-FOV truncated waveguide type metalens has also been demonstrated (*13*). However, the short focal length of the lens (717µm) did not allow direct coupling to a camera for outdoor imaging.

In this paper we demonstrate for the first time a metalens which allows outdoor imaging with natural lighting (or artificial LED lighting). The metalens, which is implemented by Huygens nanoantennas, supports a wide field-of-view of ±15˚ over a relatively broad spectral range of up to ~ 40 nm. Our findings confirm our previously published expectation that good quality outdoor imaging can be achieved with a non-chromatically corrected metalens, by judiciously choosing the aperture and spectral range (*14*). Schematic illustration of the metalens and its application for outdoor imaging is shown in fig. 1(a).

**Design**

The optical design of our metalens was performed using commercial optical design software (Zemax OpticStudio, Zemax LLC), combined with full-wave finite-difference-time domain (FDTD) simulations to determine the amplitude and phase response of the nanoantennas. The design concept is based on an aperture stop located at the front focal plane of the lens, which results in a telecentric design (chief ray exits parallel to optical axis), with good off-axis aberration correction (*13*, *15*). The layout of the optical system is shown in fig. 1(b). The front aperture is 1.35mm in diameter, and the focal length of the metalens is 3.36mm (F/2.5). The design supports field angles of up to 40˚, with near diffraction limited performance, using a quadratic diffractive phase function, as described by equation 1.

$$\phi(r) = ar^2 \;,\; a = -1098.2 \text{ mm}^{-2} \qquad (1)$$

The main performance criterion for imaging lenses is the modulation transfer function (MTF), which describes the resolution of the lens. The MTF gives the modulation (i.e. contrast) attenuation factor for each spatial frequency, as a result of the lens blur spot. The function describing the blur spot is called the point-spread-function (PSF) and is a two-dimensional function of the transverse horizontal (x) and vertical (y) image plane coordinates. Integration of the PSF in the horizontal and vertical directions yields the vertical and horizontal one-dimensional line-spread-functions (LSF), respectively. The MTF is the Fourier transform of the LSFs, so we have two MTFs, horizontal and vertical (*16*).

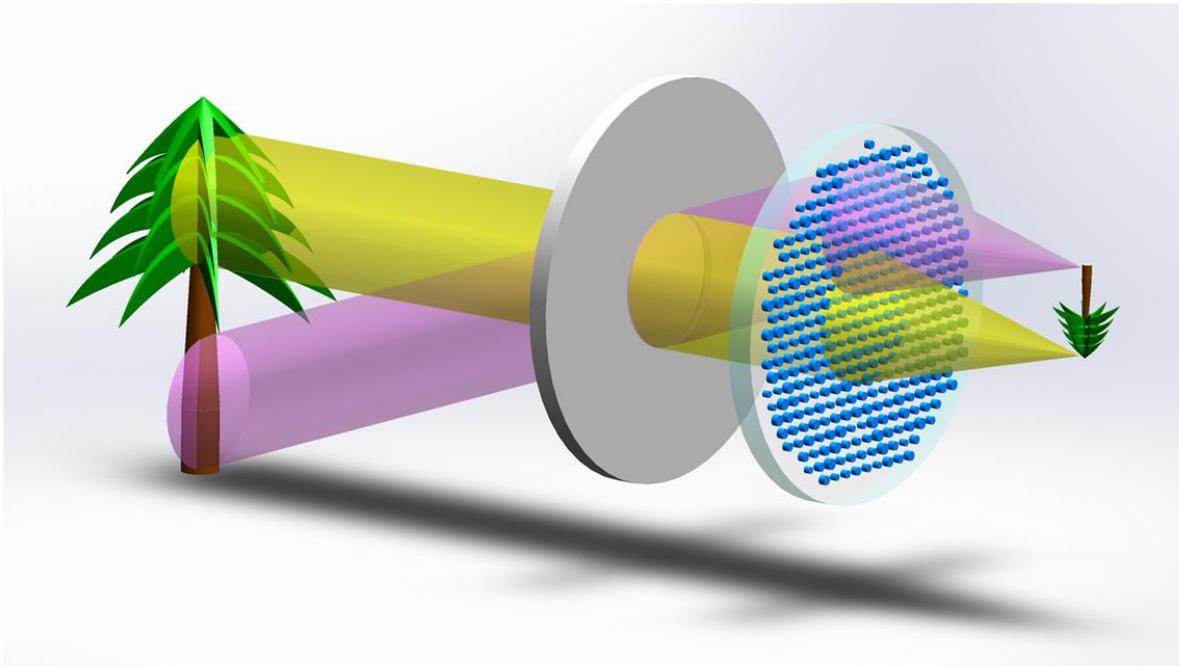

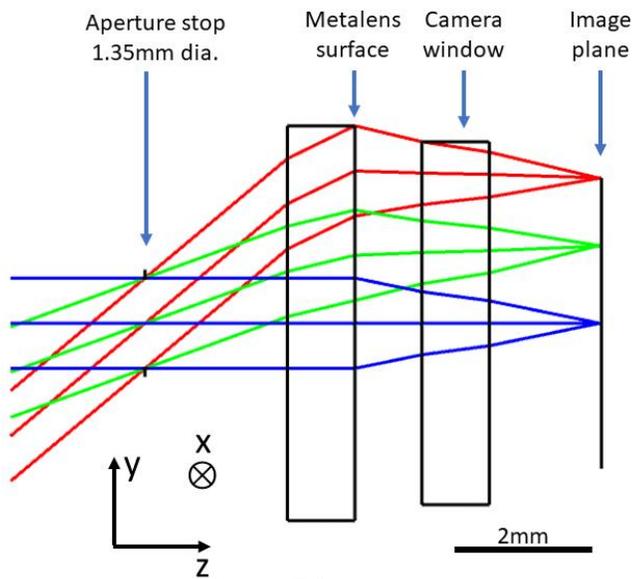
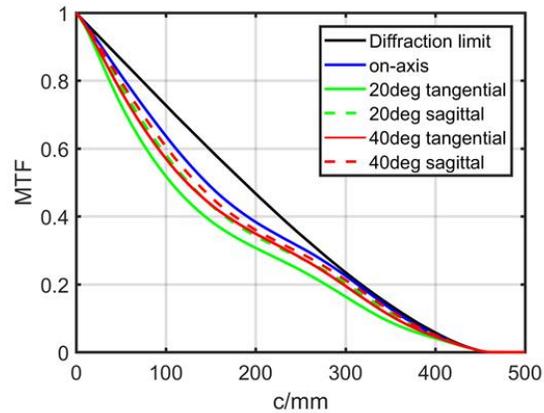

**Fig. 1. Metalens optical layout and performance.** (a) Schematic illustration of imaging with Huygens metalens. (b) Optical layout. Ray fans drawn are for 0˚, 20˚, and 40˚ incidence. (c) Nominal MTF for monochromatic 850nm illumination.

The MTF of the nominal metalens design, for monochromatic illumination at 850 nm, is shown in fig. 1(c). For the on-axis image point the nominal horizontal and vertical MTFs are identical, because of symmetry, but for the off-axis points there are two MTF graphs. The direction of the off-axis excursion is called the tangential direction. In our case this is defined the y-axis direction, as denoted in fig. 1(b). The perpendicular direction is called the sagittal direction, which is the x-axis direction in our case.

In addition to MTF, it is also critical for an imaging lens to provide illumination which is both strong enough and reasonably uniform over the field-of-view. Since our lens is designed to operate with broad-spectrum illumination, we must allow a sufficiently wide spectral range of light to pass through to obtain a good illumination level. Unfortunately, the wider the spectral range, the larger the chromatic blurring will be, resulting in a degradation of the monochromatic MTF shown in fig. 1(c) to the polychromatic MTFs that will be shown later (fig. 4(a-c)). This tradeoff between resolution and illumination signal was explored from a theoretical point-of-view in our previous paper (*14*). The results shown in this paper confirm our expectation that at outdoor illumination levels, one can obtain good signal and resolution by using moderate spectral bands of up to 40nm.

An additional performance parameter, not accounted for by MTF, is the geometrical distortion, which is a distortion of the shape of the imaged objects, without affecting the image resolution. The relative distortion is defined as the ratio between the shift in position of an image point relative to the absolute ideal position. This type of design exhibits negative ('barrel') distortion, reaching 23% at 40˚ FOV, but only 3.4% at 15˚ FOV, as shown in fig. S1(a). As a rule of thumb, a distortion of up to 10% is not disturbing to a standard viewer.

To reduce the fabrication effort, the lens aperture diameter was limited to 2mm. This resulted in blocking of some rays (which optical designers call 'vignetting') at off-axis incidence angles larger than 8˚, causing a gradual drop-off of illumination as shown in fig. S1(b). However, at incidence angles up to 15˚ we still have above 65% relative illumination (this is without considering the Huygens antenna response to different incidence angles, which will be discussed in the Efficiency section).

To implement the phase shifts required for the diffractive phase function, while maintaining high transmission, we used Huygens nanoantennas. The nanoantenna simulation was performed using commercial 3D FDTD software (Lumerical FDTD Solutions). The antennas are made of amorphous silicon on a glass substrate and are covered with a thin layer of PMMA (~300nm thick). The lattice period was chosen to be 500nm. This period was chosen considering sub-wavelength and phase sampling requirements (see details in Supplementary), in addition to antenna coupling, which occurs at smaller periods, and interferes with achieving the Kerker condition (*17*).

To find the optimal antenna dimensions, we performed a numerical scan over the antenna radius and height, while monitoring the transmission and phase of a periodic antenna array. It turned out that for a hexagonal lattice with a period of 500nm, the optimal antenna height (where the electric and magnetic dipole resonances overlap at a wavelength of 850nm) is 140nm. The transmission and phase response are shown in fig. 2.

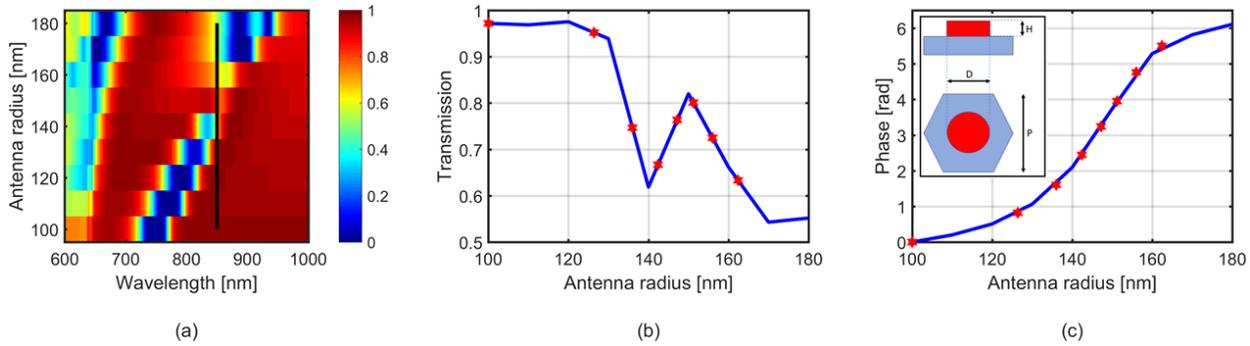

**Fig. 2. Response of Huygens aSi antenna array on glass.** (a) Transmission as a function of wavelength and antenna radius. Black vertical line is the section along which graphs b and c are drawn. (b) Transmission at 850nm as a function of antenna radius. (c) Phase at 850nm as a function of antenna radius. Red markers are at the location of the 8 antenna radii used in our metalens. Inset: Nano-antenna unit cell design. P=500nm, H=140nm, D is in the range of 200-328nm.

To implement the desired quadratic phase function, eight discrete antenna radii were chosen, spanning a range of 100-164nm, such that the phase shifts are equally spaced over the $2\pi$ range. It can be seen from fig. 2(b) that transmission higher than 60% is maintained for the full range of radii. The parabolic phase function extracted from the above mentioned Zemax design was used to determine which of the 8 values of antenna radii should be placed at each transverse location across the lens aperture. The metalens graphic layout together with optical and SEM images of the manufactured metalens are shown in fig. 3.

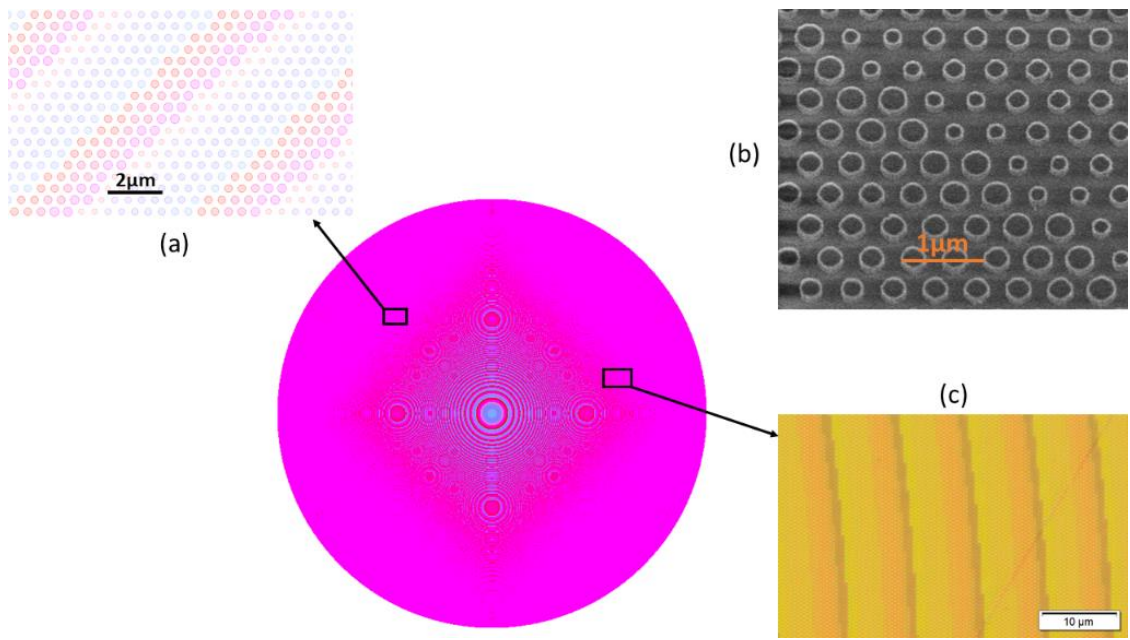

**Fig. 3. Metalens design pattern for e-beam lithography.** Each color represents a different antenna radius. The bulls-eye pattern in the center of the metalens are the Fresnel zones. Other bulls-eye patterns are aliasing artifacts. Insets: (a) Zoom in on antenna pattern. (b) SEM image of metalens section. (c) Optical microscope image of metalens section.

# Results
## Resolution

The MTF of the lens was measured on-axis and off-axis at several spectral widths, using the setup described in the Methods section. The measurement results are shown in fig. 4, in comparison to the simulated results. To the best of our knowledge this is the first time that a polychromatic MTF measurement of a metalens has been performed. The MTF simulation results are affected most by axial and lateral chromatic aberration, and vignetting effects. The spectral weights used in our simulation are based on the measured spectrum of the light source used in our test setup with different spectral filters, and spectral responsivity of the camera. The spectral contribution of the setup optics and of the metalens itself (which is the first order metalens efficiency shown later in fig. 5(a)) were neglected, since they are quite flat over the relevant spectral range.

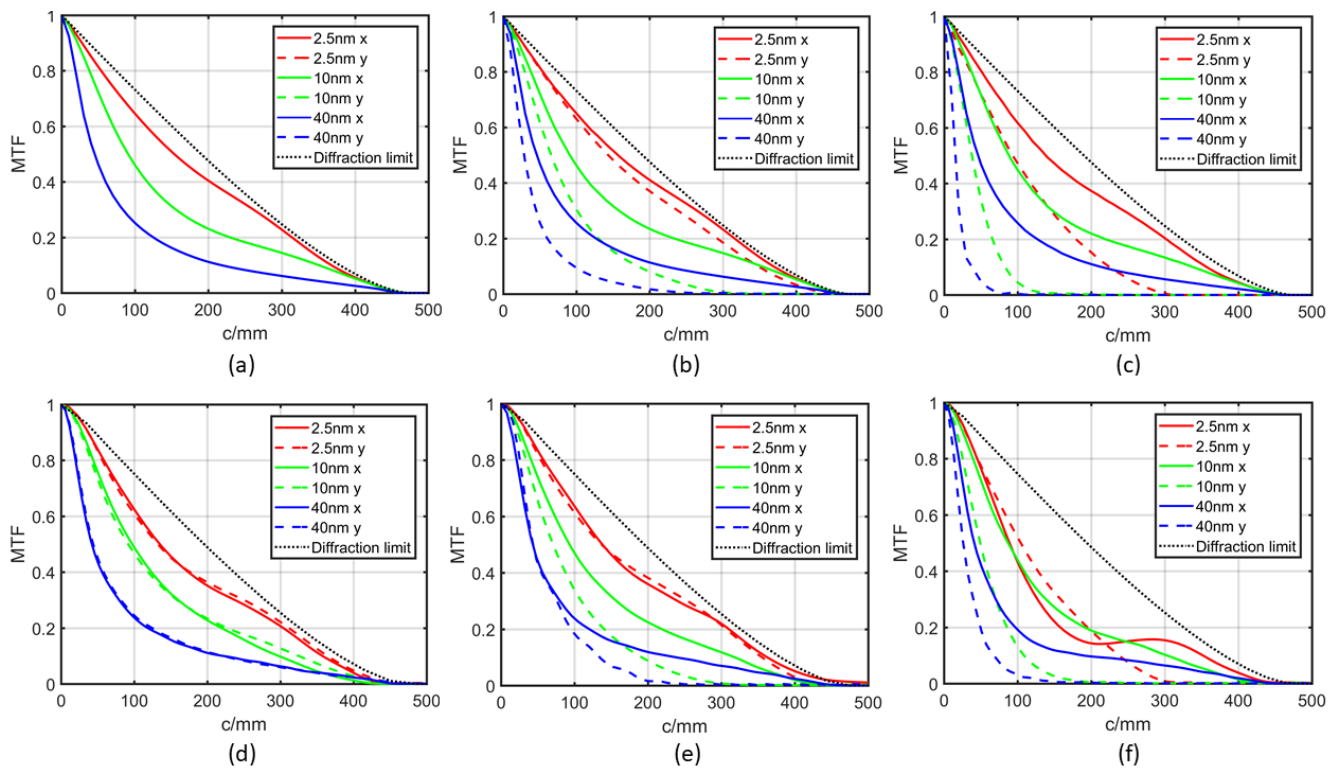

**Fig. 4. Polychromatic MTF measured vs. simulated results.** (a), (b) and (c) are simulated results for on-axis, 0.4mm off-axis, and 0.8mm off-axis (in the image plane), respectively. (d), (e), and (f) are the corresponding measured results. In each graph results for three spectral widths, 2.5nm, 10nm and 40nm are presented, and compared to the theoretical diffraction limit. The y-axis is the tangential direction (the direction in which we went off-axis).

By comparing fig. 4(a) and (d) we see an excellent match between the measured and simulated results on-axis. Note that there is only one on-axis simulated graph for each spectral width in fig. 4(a), because of symmetry, while for the measured results of fig. 4(d) there are two graphs, since in reality tolerances and noise break the symmetry even on-axis.

The measured result for 2.5nm spectral width is slightly lower than the simulation. This is most likely caused by the inaccuracy in centration of the stop relative to the metalens. For the off-axis, comparing fig. 4(b) vs. (e) and (c) vs. (f), we see a good match between theory and experiment for the y (tangential) axis. Interestingly, the measured results are slightly better than simulated, probably because we clipped the distant edges of the PSF during measurement. In the x-axis (sagittal) direction the measured results match the theory very well, except for the 2.5nm spectral width. Here the measured results are significantly worse than expected, again most likely due to the decentration tolerance between the stop and the metalens, which is more sensitive off-axis than on-axis. The effect of vignetting is felt mostly at the 0.8mm off-axis point with 2.5nm spectral width, in the y-direction. In this case the performance is diffraction limited, but the diffraction limit is lower, with a cutoff at about 300c/mm instead of 450c/mm.

### Efficiency

Next, we measure the efficiency of the metalens, i.e. what percent of the incident light gets to the first order focal point? The measurement was performed using the same setup used to measure MTF, at several wavelengths – for details see the Methods section. The results are shown in fig. 5 and are compared to expected diffraction efficiency values based on simulation. The simulation was performed in Matlab, using the transmission and phase values which were calculated in Lumerical for each of the 8 radii of antennas (for the off-axis points a beam incident at an angle was used, together with Bloch boundary conditions, for 850nm wavelength). To simplify, we did not use the lens itself for the diffraction efficiency calculation, but rather we used a blazed grating. The efficiencies in both cases are similar since the lens can be viewed locally as a grating with radially varying period (*18*). In our FDTD simulations we used the nominal antenna radii, since to the accuracy of our SEM measurement, the radii came out close to nominal. In the simulations, we used the antenna height value of 135nm, measured by surface profilometer, rather than the nominal 140nm.

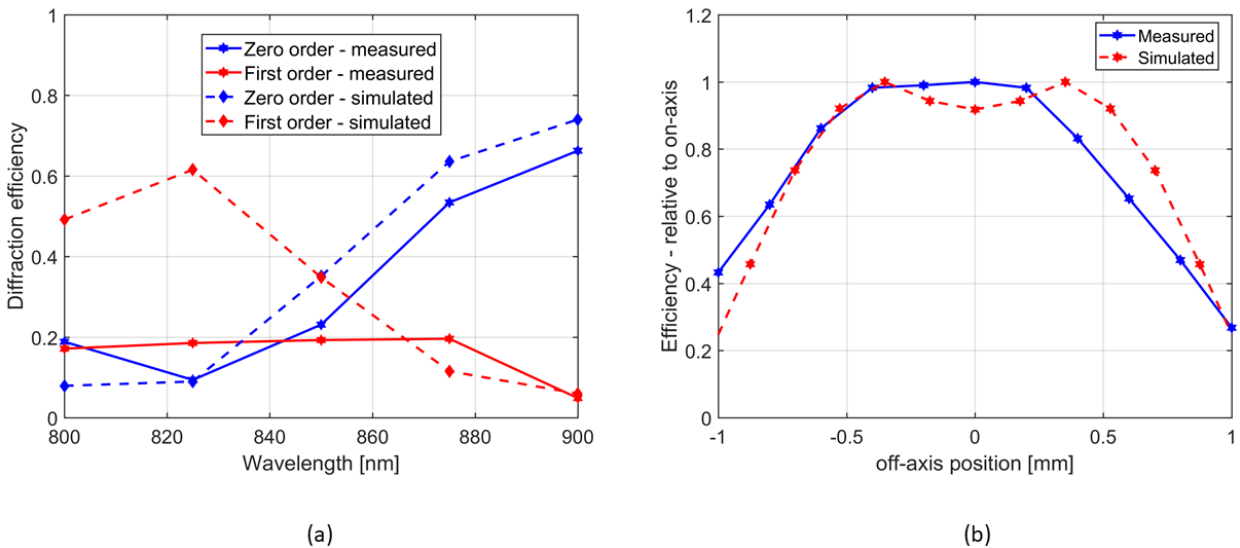

**Fig. 5. Metalens diffraction efficiency.** Measured results compared to simulation. (a) On-axis, first and zero order. (b) Off-axis, first order. 1mm off-axis in the image plane corresponds to a field angle of 17.3˚.

From the simulation results shown in fig. 5(a), a shift of the maximum efficiency wavelength from the nominal design value of 850 nm to 825 nm is observed (peak of red-dashed line). This shift is the result of the slight deviation in antenna height as mentioned above. The measured first order diffraction efficiency is much lower than the simulated values, reaching a maximum efficiency of about 20% as compared with the simulated value of 60%. Indeed, the simulation is expected to be slightly optimistic, since we did not account for the reduced number of phase steps near the edge of the metalens, and for the spatial phase sampling errors which become significant at the smaller zone widths. However, the contribution of these effects is not expected to be very significant, as the minimal diffractive period of the metalens is 2.86µm, still supporting more than 5 phase samples in the smallest period (*19*). In addition, there is some inaccuracy due to use of the scalar approximation in the simulation. Another possible contributor is the difference between the FDTD simulations and the experimental results regarding the interaction between antennas. This is because the simulation assumes a periodic structure, whereas in reality neighboring antennas may have a different radius. Furthermore, there is about 8% of power loss due to Fresnel reflection. All this being said, we anticipate that the reduced efficiency is caused mostly by errors in the antenna phase response resulting from manufacturing tolerances (errors in disk circularity, diameter, height, verticality of sidewalls etc.). This is a subject for further investigations. However, no fundamental limitations prevent the achievement of efficiency near the simulated value.

The off-axis efficiency simulation, shown in fig. 5(b), accounts for vignetting (caused by the 2mm metalens diameter) which we will call 'geometrical efficiency', and also for the variation in diffraction efficiency caused by the change in antenna response (transmission and phase) as a function of incidence angle. There is a good match between measured and simulated results. The dip in simulated efficiency at normal incidence stems from using the experimentally observed height of 135 nm rather than the optimal 140 nm. This causes optimum efficiency to be obtained slightly off-axis. This effect is not evident in the measured results, probably because of manufacturing tolerances. In addition, the measured efficiency is slightly asymmetric with respect to the off-axis position (shifted to the left). This is probably because of the slight mechanical decenter of the iris with respect to the metalens.

### Outdoor imaging

To obtain a qualitative assessment of the imaging quality of the metalens, we adapted it to a digital monochrome video camera (Thorlabs DCC1545M). In fig. 6 an outdoor picture of our Israeli research group, taken with our metalens, is shown. The scene was illuminated by natural sunlight, with a band-pass filter (Thorlabs FB800-10, central wavelength 800nm, 10nm FWHM) incorporated in front of the metalens. No post processing was performed on the image, except for cropping of the dark areas in the periphery. This is the first time ever that a metalens is used for imaging in outdoor environment.

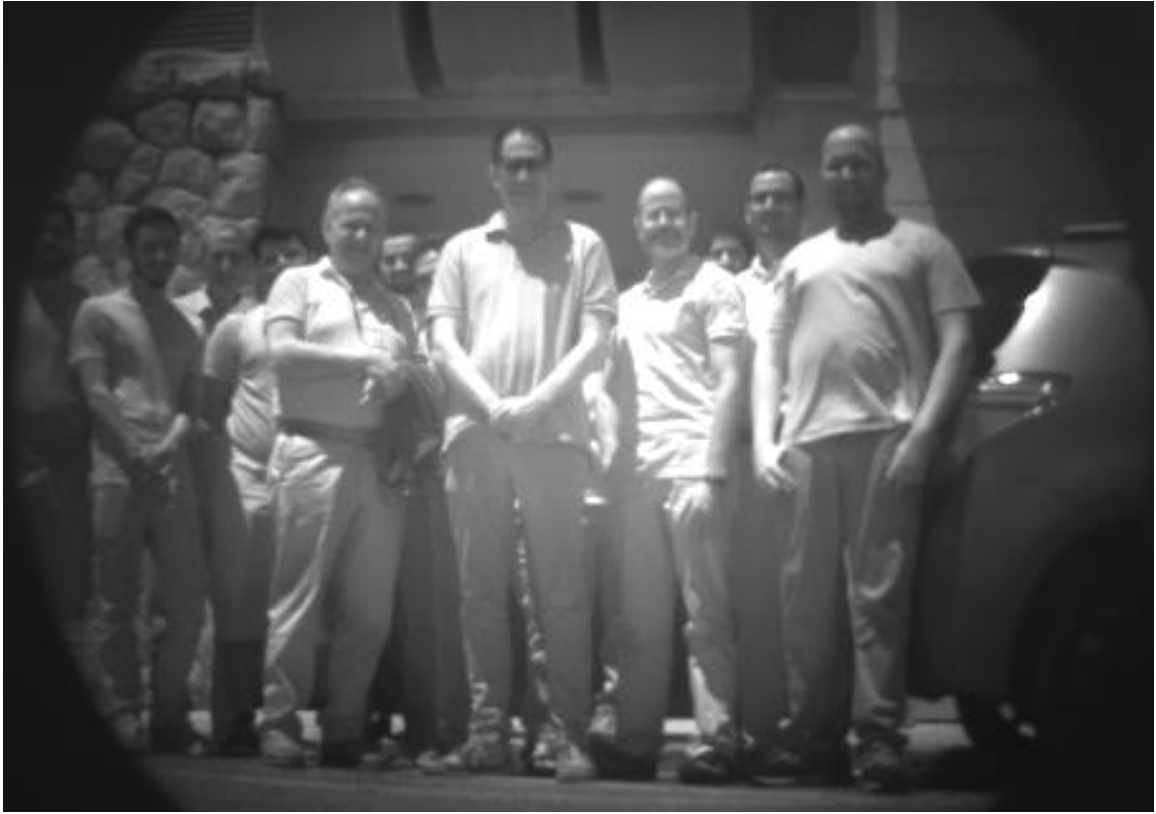

**Fig. 6. Metalens outdoor image.** NanoOpto group members in Hebrew University

**Discussion**

The advantage of the Huygens metalens over previously studied truncated waveguide or geometric phase metalenses (2, 3) is the smaller antenna aspect ratio, which simplifies the fabrication process. Based on our design the antenna height is 140nm and the smallest diameter is 200nm, so the highest aspect ratio is about 0.7. This is about an order of magnitude lower than the aspect ratio needed for the implementation of geometrical phase or truncated waveguide nanoantennas.

Alongside with the high merits of the Huygens type metalens, the Huygens type nanoantenna based metalens is more prone to angular, wavelength, and geometrical deviations. This is a result of the fact that the Huygens concept is based on a resonant effect, which naturally has a narrower spectral response as compared to the other, non-resonant approaches. We have found that while this high sensitivity degrades the efficiency of the metalens significantly, it has a marginal effect on the obtainable resolution.

The sensitivity of efficiency to manufacturing tolerances is inherent in the Huygens design, and can be overcome by an improved and more accurate fabrication process. The sensitivity of efficiency to wavelength was found to be moderate, allowing for about 40 nm bandwidth to be used, based on Fig. 5(a) (first order simulated result). In the case of a chromatic lens (i.e. not corrected for chromatic aberration) we will be limited to a moderate bandwidth, based on the resolution-illumination trade-off. An analysis of the optimal bandwidth to use for a given metalens is beyond the scope of this paper. Such a discussion can be found in our previous paper (*14*). The sensitivity to angle of incidence limits the FOV of a standard Huygens metalens to a maximum of about 10˚-20˚, as shown in Fig. 5(b).

Current metalens research is focusing on correcting the chromatic aberration of metalenses and providing tunability of the metalens focal length. Chromatic abberation can be corrected by using dispersion engineered antennas (*20*, *21*), but this works only for a narrow spectral-band or Fresnel number (which is proportional to the NA-aperture product) (*22*). Other options are transverse (*23*, *24*) or longitudinal (*25*) multiplexing of metalenses. In all cases the chromatic correction comes at the expense of efficiency and manufacturability. Several methods have been demonstrated for tuning metalens focal length – some of them external to the metalens (*12*, *26*), and some more integral to the lens (*27*–*29*).

Finally, the potential impact of optical metalenses is currently subject to discussions. They are being compared with some pioneering previous works implementing surface relief diffraction lenses. Indeed, what reason is there to choose a metalens type diffractive lens over a surface relief diffractive lens? In (*30*, *31*) it is argued that the only advantage of a metasurface over a diffractive surface is if one wants to manipulate the polarization. This would mean that conventional metalenses, that manipulate only the phase of the light, are redundant. Here, we would like to highlight several other potential advantages: (a) The metalens can accomplish the same task as the diffractive using a binary structure instead of multilevel or kinoform, so it is easier to manufacture. (b) Metalenses have less shading effect for small diffractive periods (*32*) (which is very important for high-power lenses, that necessarily have small periods near the edge). (c) Metalenses allow dispersion control of the phase, thus allowing control of spectral diffraction efficiency and chromatic aberration (*33*, *34*). It is further argued that the aspect ratio of the nanoantennas is higher than that of an equivalent diffractive type lens. While this may be the case for truncated waveguide and geometric phase type metalenses (depending on the NA of the lens, since for a diffractive element the aspect ratio increases for larger NA), the Huygens type metalens presented in this paper is based on nanoantennas with low aspect ratios. As mentioned above, the highest aspect ratio of our metalens is about 0.7, whereas the diffractive lenses discussed in (*30*, *31*) report an aspect ratio of about 1. To summarize, in our opinion surface relief diffractive elements and the various types of metasurface diffractive elements each have their advantages and disadvantages. The choice of which type of diffractive element to use is application specific and depends upon manufacturing capabilities and performance requirements.

**Materials and Methods**

The MTF simulation shown in fig. 4(a-c) was performed using Matlab in conjunction with Zemax. The simulation could not be performed simply in Zemax, because of the large number of wavelengths required to obtain accurate MTF results for a diffractive lens with large chromatic aberration (Zemax supports only up to 24 wavelengths, while we used 85, 160 and 453 wavelengths for spectral widths 2.5, 10 and 40nm respectively). Therefore, a ZPL macro was written in Zemax to calculate the optical transfer function (OTF) at each single wavelength and output the data to a text file. Matlab was then used to sum the OTFs of all the wavelengths, with appropriate spectral weighting, to obtain the total OTF. The absolute value of the OTF was then taken to obtain the MTF.

In our previous paper (*14*) we simulated the MTF of a metalens assuming a top-hat shaped PSF at each wavelength, resulting from chromatic defocus, which is a geometrical optics approximation. While the geometrical optics approximation is accurate for large defocus, such that the distance from the focal spot to the image plane can be considered 'far-field', for our case this is not accurate, since the chromatic defocus is relatively small. For the simulations in this paper we therefore used a physical optics calculation based on the Fraunhofer approximation, as performed by the 'FFT MTF' simulation in Zemax. We verified that the defocus spots are small enough in our case so that the Fraunhofer approximation is accurate – see Supplementary.

Our metalens was fabricated on a 1mm thick glass substrate using electron beam lithography. A 135nm thick layer of amorphous silicon was deposited on the glass using PECVD (Oxford Instruments Plasmalab System 100). The substrate was then spin coated with CSAR electron beam resist, and thermal evaporation of a 20nm thick aluminum layer for de-charging was performed. Following electron beam writing, the CSAR was developed using AR 600-546, and a 50nm thick layer of aluminum was deposited by electron beam evaporation, for use as a hard-mask. Lift-off was then performed using AR 600-71 solvent. Following lift-off, the sample was etched using RIE (Corial 200I), and the aluminum mask was removed using aluminum etch solvent (J.T.Baker 80-15-3-2). The sample was then spin coated with a 300nm thick layer of PMMA for use as an index matching protective cover.

MTF and efficiency were measured using the experimental setup shown in fig. S2. The fiber coupled light source is used as an input to the collimator, to produce a high-quality collimated beam. The collimated beam is used as the input to the metalens. A mechanical stop is placed in front of the metalens, at the nominal stop position. Note that the mechanical stop is centered with respect to the metalens, but when measuring off-axis the beam wanders to the side of the metalens, since the stop is removed from the metalens, and is located at its' front focal plane. The metalens focuses the light to a point, the intensity profile of which is our PSF. The PSF is then measured using a 50X microscope.

The efficiency was measured by integrating the camera pixel values for several situations, shown in fig. S3: (a) Same as MTF measurement – gives the power in first diffraction order. (b) 50X objective removed – gives the power in the zero order. (c) Metalens removed, but aperture stop still in place – gives the total power, which is our reference signal.

# Supplementary Materials

## Supplementary Figures

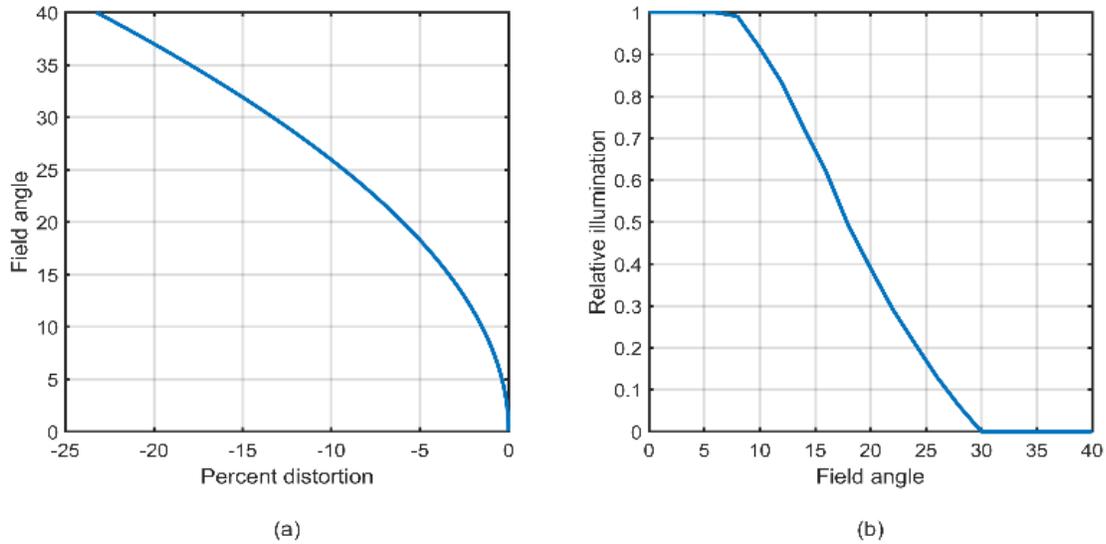

(a)  (b)

**Fig. S1. Distortion and illumination.** (a) Percent distortion as a function of field angle. (b) Relative illumination as a function of field angle, for a metalens of 2mm diameter.

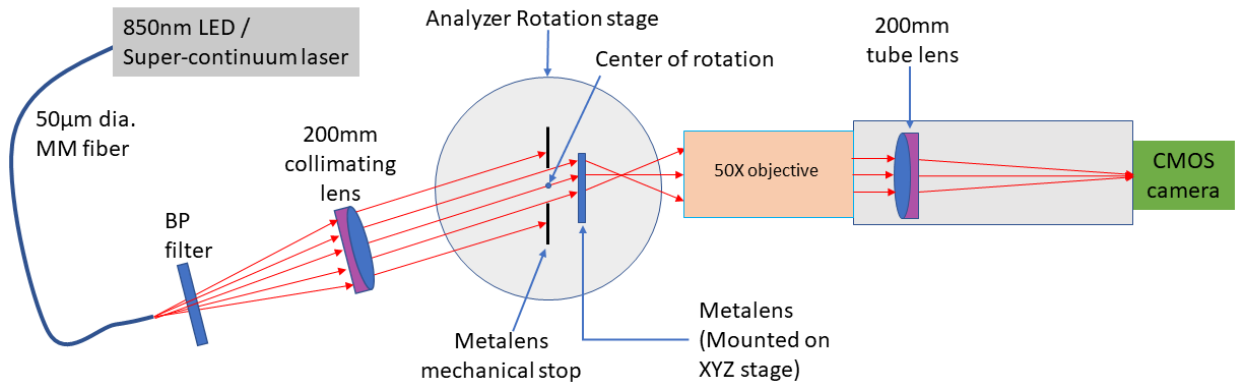

**Fig. S2.** MTF measurement setup

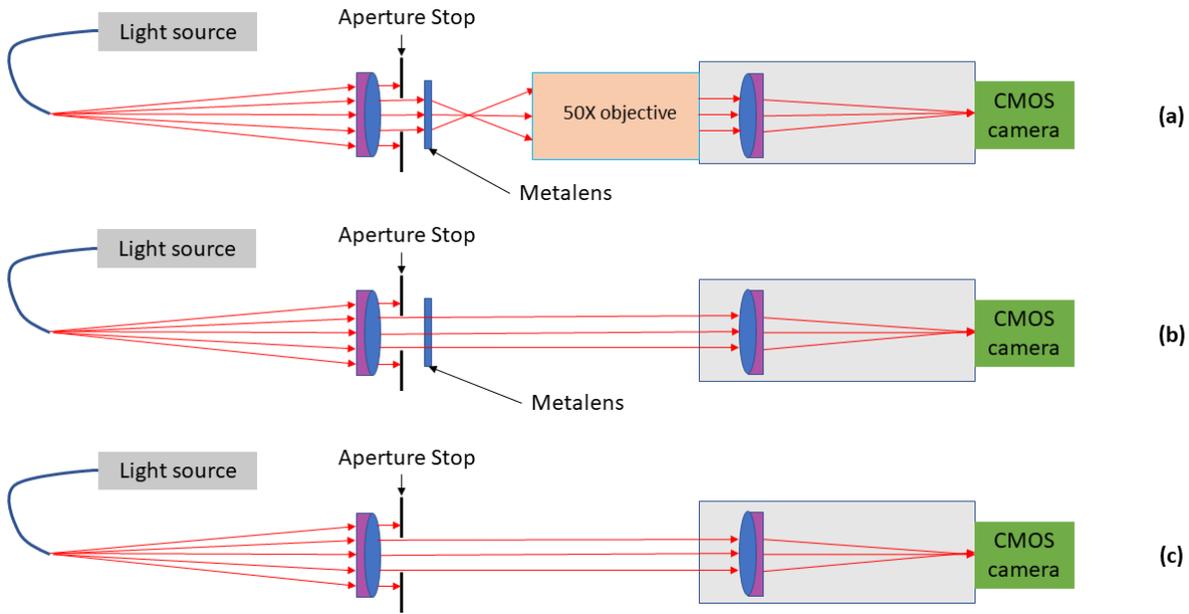

**Fig. S3** Diffraction efficiency measurement using MTF setup

# Choosing the lattice period

There are two primary factors limiting the maximum lattice period that can be used in a metasurface:

(a) The period must be sufficiently small so that only the zero diffraction order will exist (*13*).

(b) The period must be sufficiently small to allow good spatial sampling of the desired phase pattern on the lens (*35*).

Considerations (a) and (b) can be quantitatively analyzed using the grating equation (*36*):

$$(n\sin\theta - n'\sin\theta') = \frac{m\lambda}{d} \qquad (2)$$

Where $n, n'$ are the refractive indices in the incident and transmitted medium respectively, $\theta, \theta'$ are the incident and diffracted beam angles, $m$ is the order of diffraction, $\lambda$ is the wavelength, and $d$ is the grating period.

This equation can be written in alternative form in k-vector space, if we multiply both sides by $2\pi/\lambda$. In this way we obtain (*37*):

$$k_y - k'_y = mK_G \qquad (3)$$

Where $k_y$ represents the tangential component of the input wave-vector, $k'_y$ is the same for the output, and $K_G$ is the grating wavenumber, defined as $K_G = 2\pi/d$.

## Sub-wavelength criterion

The borderline situation when only the zero order will propagate, is when for one of the first orders (±1), $\theta'=\pm 90°$ (while for the other first order, there will be no solution for θ'). To simplify a bit, we will consider a situation in which $n'=n$ (can be realized by immersing the nano-antennas in a polymer, index-matched to the substrate). We substitute these requirements, and $m=1$, into equation 1, to obtain:

$$d = \frac{\lambda}{n(\sin\theta \pm 1)} \tag{4}$$

We will choose the sign of 1 to be positive, so that $d$ will be positive. If we denote the incidence angle of the beam in air, before entering the substrate, as $\phi$, we can substitute Snell's law $n\sin(\theta) = \sin\phi$ and write:

$$d = \frac{\lambda}{\sin\phi + n} \tag{5}$$

For a metalens working at infinite conjugate, $\phi$ is the field angle of the lens, and $n$ is the refractive index of the substrate.

### Phase sampling criterion

The diffraction efficiency of a multilevel phase grating is given by $\eta = \text{sinc}^2(1/N)$, where N is the number of phase levels, and the sinc function is defined as $\text{sinc}(x) = \sin(\pi x)/(\pi x)$ (5). From here we obtain an efficiency of 40.5% for two phase levels, 81% for 4 levels, and 95% for 8 levels. In the case of a metalens, we have a varying phase function period, so we must decide what our sampling criterion is. In (35) the criterion used is "Nyquist" level sampling at the edge of the metalens, i.e. 2 phase samples at the edge of the lens, where the period is smallest. Since two phase levels provide only 40.5% efficiency we preferred to use a criterion of 4 phase samples (38).

The maximum lattice period from the phase sampling point-of-view, can be found based on the phase function of the metalens. The local grating period of a first-order diffractive lens is given by(39):

$$p = \frac{2\pi}{d\phi/dr} \tag{6}$$

Where ϕ(r) is the radial phase function of the lens, which in the case of our metalens is given by eq. 1. Substituting from eq. 1 into eq. 6 we obtain eq. 7:

$$p = \frac{\pi}{ar} \tag{7}$$

To find the minimum period $p_{\min}$ we must substitute the maximum aperture radius r of the metalens. In our case, with the stop located at the lens front focal plane, this is given by eq. 9 (see fig. S4):

$$r_{\max} = f \cdot \tan\theta + r_{stop} \tag{8}$$

Where f is the lens focal length, θ the field-of-view angle, and rstop is the radius of the aperture stop, in our case 1.35/2=0.675mm.

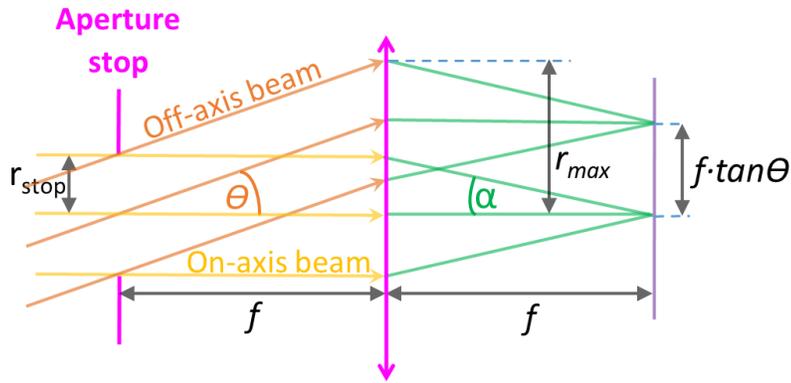

**Fig. S4** Geometry for wide-FOV metalens lattice period calculation

The required lattice period is then half of $p_{min}$, based on the Nyquist criterion.

The maximum lattice period, calculated based on the sub-wavelength and phase sampling considerations, is shown in fig. S5. The crossing point of the two graphs comes out at a field angle of 15° and lattice period of 455nm.

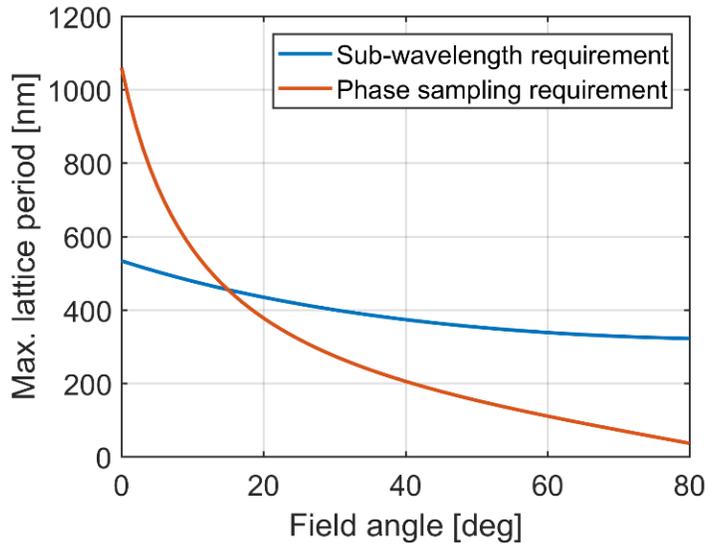

**Fig. S5** Maximum metalens lattice period as a function of field angle

Hexagonal lattice

In our metalens we used a hexagonal lattice, as shown in fig. S6. The lattice constant is p, but the effective periods in the y and x directions are $p_y = p/2$ and $p_x = p\sqrt{3}/2$. The maximum lattice period calculated above must be applied to $p_x$, the larger of the two. In our metalens we used a hexagonal lattice period $p$ of 500nm. This means that $p_x$=433nm. Based on fig. S5 we can meet this sampling criterion for field angles up to 16.25°, limited by the phase sampling criterion.

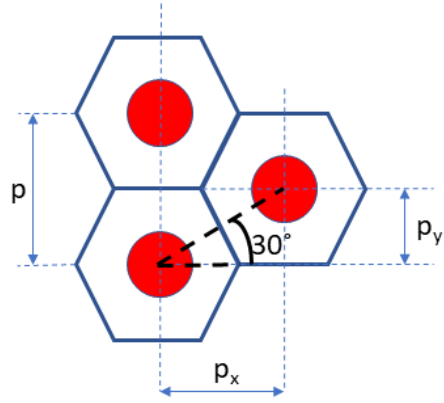

**Fig. S6** Hexagonal array effective periods

The above analysis was used to provide an estimate of the appropriate lattice constant for a given design. It gives only the upper limit for the lattice period, and not the lower limit. It was therefore followed by accurate FDTD analysis of the periodic antenna structures, and approximate scalar simulation of diffraction efficiency, as detailed in the main part of the paper.

# Fraunhofer approximation for metalens PSF

Assuming a plane wave incident on a lens, the PSF can be accurately evaluated based on the Fraunhofer approximation of the diffraction integral at the focal plane of the lens (*16*). However, in the case of our metalens the focal plane is different for each wavelength. We are evaluating the PSF of each wavelength at a common image plane, which is located somewhere in between the focal planes of the various wavelengths (at the location which gives an optimal total PSF). The question is, can we still use the Fraunhofer approximation when calculating the chromatic defocused PSFs?

The condition for validity of the Fraunhofer approximation in the case of a converging beam is given in(*40*) as:

$$x_{2\max} < 0.6\left(\lambda z^3\right)^{1/4} \qquad (9)$$

Where $x_{2\max}$ is the maximum radial spot size at the image plane, and $z$ is the propagation distance (in our case approximately equal to the focal length). If we substitute $\lambda=850nm$ and $z=f=3.36mm$ into eq. 9, we obtain the condition $x_{2\max}<250\mu m$. We can easily estimate our maximum spot size radius based on geometrical optics considerations. The metalens longitudinal chromatic aberration is given by $\Delta f = f \cdot \Delta\lambda/\lambda$, and we multiply by the tangent of the marginal ray angle (α in fig. S5) to obtain the lateral aberration (*14*). The largest spectral range we simulated was 40nm FWHM. The $\Delta\lambda$ should be half the total wavelength range, since the image plane is situated approximately in the middle. However, we also used wavelengths beyond the half-max point, so it is a good approximation to use $\Delta\lambda=40nm$. This gives us $\Delta f \approx 3.36 \cdot 40/850 \cdot 0.2 = 32\mu m$. This is far below 250µm, so we can use the Fraunhofer approximation.

**Acknowledgments**

**Funding:** The Israel Ministry of Science.

**Author contributions:**
J. E. performed the design, simulations and measurements, J.B.D provided support for the simulations and the measurements. N.M performed most of the fabrication steps of the metalens, including deposition, RIE, and complementary processes. C.Z. and A.K. where in charge of the electron beam lithography including sample preparation before, and lift-off after. U.L supervised the project. All authors contributed to discussions of the results and writing of the manuscript.

**Competing interests:** There are no competing interests.